# IceCube experience using XRootD-based Origins with GPU workflows in PNRP

*David* Schultz[1], *Igor* Sfiligoi[2], *Benedikt* Riedel[1], *Fabio* Andrijauskas[2], *Derek* Weitzel[3], and *Frank* Würthwein[2]

[1]University of Wisconsin–Madison, Madison, WI 53715, USA
[2]University of California San Diego, La Jolla, CA 92093, USA
[3]University of Nebraska-Lincoln, Lincoln, NE 68588, USA

**Abstract.** The IceCube Neutrino Observatory is a cubic kilometer neutrino telescope located at the geographic South Pole. Understanding detector systematic effects is a continuous process. This requires the Monte Carlo simulation to be updated periodically to quantify potential changes and improvements in science results with more detailed modeling of the systematic effects. IceCube's largest systematic effect comes from the optical properties of the ice the detector is embedded in. Over the last few years there have been considerable improvements in the understanding of the ice, which require a significant processing campaign to update the simulation. IceCube normally stores the results in a central storage system at the University of Wisconsin–Madison, but it ran out of disk space in 2022. The Prototype National Research Platform (PNRP) project thus offered to provide both GPU compute and storage capacity to IceCube in support of this activity. The storage access was provided via XRootD-based OSDF Origins, a first for IceCube computing. We report on the overall experience using PNRP resources, with both successes and pain points.

## 1 Introduction

The IceCube Monte Carlo simulation compute workflows have traditionally relied on GridFTP file transfers to deliver the outputs back to central storage servers located at the University of Wisconsin–Madison (UW). Unfortunately, that storage area got close to being out of disk space in late 2022, so an alternative temporary storage location was needed for some of the produced outputs.

The most computationally intensive part of the IceCube simulation workflow is a photon propagation code [1], which greatly benefits from running on GPUs. The Prototype National Research Platform (PNRP) Kubernetes cluster [2-3] is one of the largest GPU-providing compute resources for IceCube, but has never provided long-term storage areas to IceCube before. Given the storage shortage at UW, IceCube decided to investigate what PNRP had to offer.

PNRP is a globally distributed infrastructure and operates several storage areas in different parts of the globe. The available storage areas were, however, managed using the XRootD software [4] and configured as Open Science Data Federation (OSDF) Origins [5-6], which IceCube had no prior experience with. We thus had to modify the IceCube workflow to use this alternative mechanism.

## 2 Changes needed to the IceCube workflow

IceCube uses HTCondor [7] as its batch workload management system. At the time of this exercise, IceCube jobs would explicitly move data inside their jobs using a home-grown wrapper script, relying on HTCondor only for compute resource scheduling and propagation of X509-based credentials used by GridFTP.

There were two major changes required to make use of the OSDF Origins on PNRP. First, the tools used to access XRootD-based OSDF Origins are different than the ones needed to access GridFTP servers. Second, the authentication is based on SciTokens [8] instead of X509 credentials. Between the two, the changes to the IceCube wrapper scripts would have been non-trivial.

Recent versions of HTCondor have added native support for OSDF Origin endpoints, so we decided to forego the wrappers and rely on native HTCondor file transfer capabilities instead. This added some additional complexity to both the HTCondor infrastructure as well as to the IceCube's HTCondor job submission procedures, which we outline below.

### 2.1 SciToken handling

HTCondor has the capability of issuing its own SciTokens credentials, by means of its CredD process [9]. We thus configured the IceCube's HTCondor instance accordingly and worked with the PNRP OSDF team on establishing the needed trust relationship. The relevant HTCondor config section used is available in Figure 1.

```
LOCAL_CREDMON_PROVIDER_NAME = scitokens
JOB_TRANSFORM_NAMES = $(JOB_TRANSFORM_NAMES) AddSciToken
JOB_TRANSFORM_AddSciToken @=end
[ Requirements = (JobUniverse =?= 5 && ifThenElse(isUndefined(NeedsOSDF),
  False, NeedsOSDF));
  Eval_Set_OAuthServicesNeeded = strcat( "scitokens ",
  OAuthServicesNeeded ?: ""); ]
@end
# Change this to match the OSDF issuer name
LOCAL_CREDMON_ISSUER = https://chtc.cs.wisc.edu/icecube
LOCAL_CREDMON_TOKEN_AUDIENCE = ANY
# Change this for the paths your token should write into,
# relative to the IceCube root
LOCAL_CREDMON_AUTHZ_TEMPLATE = read:/production write:/production
LOCAL_CREDMON_PRIVATE_KEY = /etc/condor/.secrets/scitokens_private.pem
# Change this to match the key ID generated
LOCAL_CREDMON_KEY_ID = 7672
```

**Fig. 1.** HTCondor SciTokens config

With the proper HTCondor infrastructure in place, any job submitted to this HTCondor instance was issued a short-lived SciTokens credential, which was automatically copied to each running job and renewed as-needed by HTCondor.

**2.2 HTCondor job submission changes**

In addition to the simplification of the job wrapper script, the next major change to IceCube's workflow was the addition of input and output file locations in each job's submission description itself. This is achieved by using the *'osdf://'* prefix in the HTCondor file transfer URL [10] and by explicitly restricting jobs to resources that support this transfer type, by adding *'regexp("osdf",HasFileTransferPluginMethods)'* to the job requirements. The comparison of the original and updated workflows is outlined in Table 1.

**Table 1.** Summary comparison of the original and updated IceCube's workflow.

| Original workflow | Updated workflow |
|---|---|
| 1. IceCube framework refreshes x509 credential (if needed) | 1. Job submission to HTCondor by IceCube framework |
| 2. Job submission to HTCondor by IceCube framework | 2. HTCondor generates SciToken refresh and access tokens (if they do not exist) |
| 3. HTCondor transfers x509 credential with job | 3. HTCondor refreshes SciTokens periodically (even while job is idle) |
| 4. Job transfers input files via GridFTP | 4. HTCondor transfers SciToken access token with job |
| 5. Job runs simulation code | 5. HTCondor transfers input files using OSDF plugin and XRootD |
| 6. Job transfers output files via GridFTP | 6. Job runs simulation code |
| 7. Job completion. | 7. HTCondor refreshes SciTokens, and sends updated access token to job |
| | 8. HTCondor transfers output files using OSDF plugin and XRootD |
| | 9. Job completion. |

The definition of the file transfer location at submission time had, however, one major unforeseen side effect: we had to explicitly partition the jobs between the various OSDF Origins operated by PNRP. Since two of the Origins were co-located with large GPU clusters, located at the University of Nebraska-Lincoln (UNL) and Massachusetts Green High Performance Computing Center (MGHPCC), we forced all output from jobs running at those GPU resources to go to the co-located Origins, to both maximize throughput and minimize latency. Output from jobs running on all other resources went to the Origin at the San Diego Supercomputing Center (SDSC). This fixed partitioning added a bit of overhead to the IceCube's operations load, but it was not a showstopper.

## 3 Usage experience

Since this was one of the first PRNP workflows using OSDF storage and HTCondor-issued SciTokens, the setup took several weeks of fiddling with the details on both sides (storage and submit point) until it worked consistently.

Once finalized, it has been stable, although there were some pain points around the storage interface. For example, when there were transfer errors, it wasn't clear what had failed or why. Additionally, there seems to be an occasional delay between job submission and token generation that can result in a rejected job submission by HTCondor.

Nevertheless, the CPU and GPU computing side of PNRP was nearly flawless in execution, with very low error rates, well below what IceCube typically experiences in its regular computing setup.

### 3.1 Data produced

The setup was finalized in December 2022 and the simulation jobs ran until May 2023. During that time IceCube produced about 160 TB of data, almost equally distributed among the three sites, as shown in Figure 2.

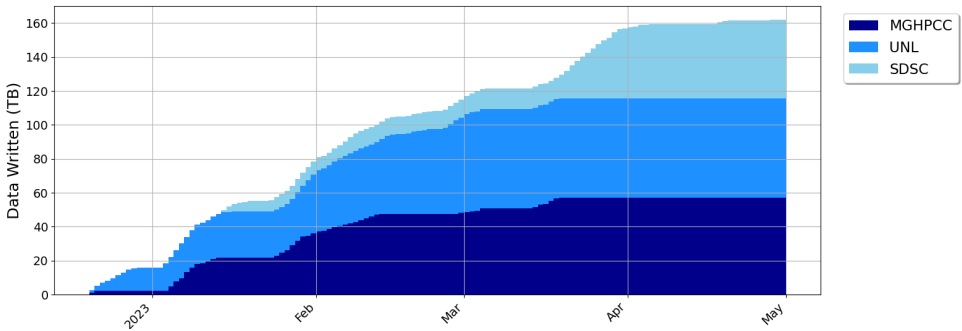

**Fig. 2.** Storage per site, cumulative.

The resources available on PNRP to IceCube were not uniform, with occasional spikes followed by periods of inactivity. Each OSDF Origin typically received less than 2 TB of data per day, but there were occasional peaks exceeding 3 TB per day, as shown in Figure 3. The detailed file transfer data are available in [11].

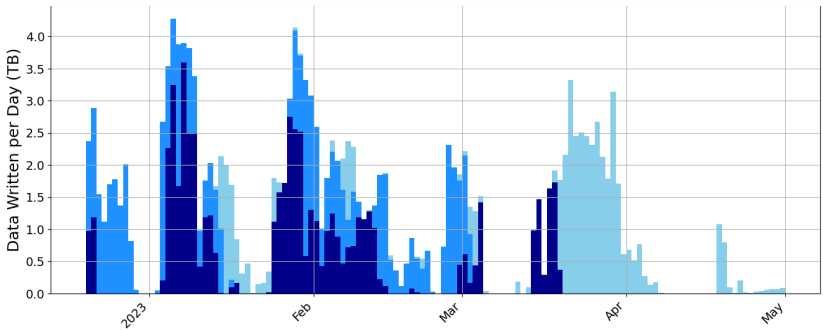

**Fig. 3.** Storage per site, binned by day.

## 4 IceCube science motivation

The IceCube Neutrino Observatory [12] is the world's premier facility to detect neutrinos with energies above 1 TeV and an essential part of multi-messenger astrophysics. IceCube is composed of 5160 digital optical modules (DOMs) buried deep in glacial ice at the geographical south pole. Neutrinos that interact close to or inside of IceCube produce secondary particles, often a muon. Such secondary particles produce Cherenkov (blue as seen by humans) light as they travel through the highly transparent ice. Cherenkov photons detected by DOMs can be used to reconstruct the direction and energy of the parent neutrino.

Since the detector is built into a naturally existing medium, i.e., glacial ice, there was a priori only limited information regarding the optical properties of the detector, so a lot of Monte Carlo simulation data is needed to properly calibrate the employed instruments. The optical properties of the glacial ice greatly affect the pointing resolution of IceCube. Improving the pointing resolution has two effects in this case: greater chance to detect astrophysical neutrinos and better information sent to the community. While IceCube can detect all flavors and interaction channels of neutrinos, about two-thirds of the flux reaching IceCube will generate a detection pattern with a large angular error. And this angular error is mostly driven by systematic effects. Similarly, different optical models have a great effect on the reconstructed location of an event in the sky. The comparatively minute field of view of partner observatories and telescopes requires IceCube to provide as accurate as information as possible.

Moreover, understanding detector systematic effects is a continuous process. This requires the Monte Carlo simulation data to be updated periodically to quantify potential changes and improvements in science results with more detailed modeling of the systematic effects. Over the last few years there have been considerable improvements in the understanding of the ice, which require a significant processing campaign to update the simulation. The workflow run on PNRP was part of one such processing campaign.

## 5 Summary and conclusions

The storage provided by the PNRP project was a great asset for IceCube, allowing them to continue their simulation compute when their native storage areas become temporarily full. The new services did, however, come with an initial startup cost, as IceCube has never used XRootD-based OSDF Origins before.

To minimize development work, IceCube switched from embedded, in-job file transfers to explicit, workload-managed, HTCondor file transfers. This slightly changed their operational model, with the major downside being the need for explicit, submit-time partitioning of the jobs between the target storage locations. Nevertheless, the resulting setup was very stable, with near-perfect stability in compute resource usage, well above what IceCube typically experiences in its regular computing setup. In the future, we will investigate transparent storage location selection in order to avoid this manual partitioning.

The PNRP storage service performance was excellent, with only a few occasional, transient transfer errors. Each OSDF Origin typically received less than 2 TB of data per day, with occasional peaks exceeding 3 TB per day. In total, IceCube produced about 160 TB of data over the course of about 5 months.

Overall, IceCube is very satisfied with the experience and will consider the PNRP OSDF Origins in the future, if additional temporary storage is needed again.

## Acknowledgments

This work was partially funded by the U.S. National Science Foundation (NSF) under grants OAC-1826967, OAC-1541349, OPP-2042807, OAC-2030508, OAC-2112167 and CNS-1730158.

## References

1. D. Chirki, *Photon tracking with GPUs in IceCube*. Nucl. Inst. and Meth. in Phys. Res. Sec. A, Vol. 725, 141-143. (2013) https://doi.org/10.1016/j.nima.2012.11.170
2. *NRP*, Accessed August 2023. https://www.sdsc.edu/services/hpc/nrp/index.html
3. *National Research Platform*, Accessed August 2023. https://nationalresearchplatform.org
4. L. Bauerdick et al., *Using Xrootd to Federate Regional Storage,* J. Phys.: Conf. Ser. 396 042009 (2012) https://doi.org/10.1088/1742-6596/396/4/042009
5. D. Weitzel et.al., *Data Access for LIGO on the OSG*. In Proceedings of the Practice and Experience in Advanced Research Computing 2017 on Sustainability, Success and Impact (PEARC17). Association for Computing Machinery, New York, NY, USA, Article 24, 1–6. (2017) https://doi.org/10.1145/3093338.3093363
6. *CVMFS – Nautilus Documentation*, Accessed August 2023. https://ucsd-prp.gitlab.io/userdocs/storage/cvmfs/
7. D. Thain, T. Tannenbaum, and M. Livny, *Distributed computing in practice: the Condor experience*. Concurrency Computat.: Pract. Exper., 17: 323-356. (2005) https://doi.org/10.1002/cpe.938
8. A. Withers et al., *SciTokens: Capability-Based Secure Access to Remote Scientific Data*. In Proceedings of the Practice and Experience on Advanced Research Computing (PEARC '18). Association for Computing Machinery, New York, NY, USA, Article 24, 1–8. (2018) https://doi.org/10.1145/3219104.3219135
9. A. Withers et al., *SciTokens: Demonstrating Capability-Based Access to Remote Scientific Data using HTCondor*. In Proceedings of the Practice and Experience in Advanced Research Computing on Rise of the Machines (learning) (PEARC '19). Association for Computing Machinery, New York, NY, USA, Article 118, 1–4. (2019) https://doi.org/10.1145/3332186.3333258

10. *Submitting jobs without a shared file system: HTCondor 's File Transfer Mechanism*. Accessed August 2023. https://htcondor.readthedocs.io/en/v10_0/users-manual/file-transfer.html
11. *GitHub raw data: CHEP23-xrootd-NRP-data*, Accessed August 2023. https://github.com/WIPACrepo/CHEP23-xrootd-NRP-data
12. M. G. Aartsen et al., J. Inst. **12** P03012 (2017) https://doi.org/10.1088/1748-0221/12/03/P03012